# Observation of square-like moiré lattice and quasicrystalline order in twisted rock-salt nitrides

**Authors:**
Dongke Rong,[1,2,†] Qinghua Zhang,[1,2,†,*] Ting Cui,[1,2] Qianying Wang,[1,2] Hongyun Ji,[1,2] Axin Xie,[1,2] Songhee Choi,[1] Qiao Jin,[1,3] Chen Ge,[1,2,4] Can Wang,[1,2] Shanmin Wang,[4] Kuijuan Jin,[1,2,*] and Er-Jia Guo[1,2,*]

**Affiliations:**
[1] Beijing National Laboratory for Condensed Matter Physics and Institute of Physics, Chinese Academy of Sciences, Beijing 100190, China
[2] Department of Physics & Center of Materials Science and Optoelectronics Engineering, University of Chinese Academy of Sciences, Beijing 100049, China
[3] School of information Science and Engineering, Shandong Agricultural University, Taian 271018, China
[4] Department of Physics, Southern University of Science and Technology, Shenzhen 518055, China
†These authors contribute equally to the manuscript.
*Corresponding author Emails: zqh@iphy.ac.cn, kjjin@iphy.ac.cn and ejguo@iphy.ac.cn

**Abstract**
Twistronics, which exploits moiré modulation of lattice and electronic structures in twisted bilayers, has emerged as a powerful approach to engineer novel quantum states. Recent efforts have expanded beyond two-dimensional van der Waals (vdWs) crystals to more complex, strongly correlated materials, where interfacial moiré effects can dominate physical properties. Here we demonstrate a generalizable route to fabricate twisted bilayers of transition metal nitrides (TMNs) with vdWs–like interfaces, using freestanding CrN membranes as a model system. Twisted bilayer CrN (tCrN) is realized by employing cubic alkaline-earth metal monoxides as sacrificial layers, enabling the assembly of clean, controllable interfaces. Electron ptychography reveals well-defined, periodic square moiré superlattices in tCrN. For a twist angle of 16.3°, we identify a nearly commensurate moiré lattice with coincident Cr columns, whereas at 45° we uncover localized octagonal quasicrystalline order with clear self-similarity. These results establish a practical platform for twisted TMNs and open avenues to explore moiré-induced atomic configurations and emergent correlated phenomena in nitride-based heterostructures.

**Main text**

Van der Waals (vdW) interfaces without valence bonding in twisted 2D crystals can profoundly modulate the atomic structures and related physical properties. In the magic-angle twisted bilayer graphene (MATBG) with a twisted angle approximately 1.1°, the moiré potential reconstructs the mini-Brillouin Zone and generates flat bands, leading to tunable correlated electron states such as Mott insulator and superconductivity[1-3]. Besides, the topological phases, electronic lattices, and moiré excitons have been reported in MATBG and its derivative systems, including twisted trilayer graphene and twisted transition metal dichalcogenides (TMDs) [4-13]. The commensurability plays a key role in moiré modulation: the transitional symmetry gives rise to long-range moiré potential, and the separated bilayers are strongly coupled via phase coherence. Spontaneous atomic reconstruction can locally realize energetically favorable commensurate stacking orders even within an overall incommensurate structure[14-16]. Moreover, recent studies have shown that moiré modulation can strongly alter the electronic structures in incommensurate twisted systems. The incommensurate 30° twisted 2D crystals, such as 30°-tBLG and 30°-$tWSe_2$, where the separated bilayers are strongly coupled via Umklapp scattering mechanism[17-19]. The moiré modulation on the interfacial structures serves as a new degree of freedom to investigate novel properties in twisted bilayer systems.

Strongly correlated electronic materials are characterized by the coupling of multiple order parameters, including charge, spin, orbit, and lattice degrees of freedom, which gives rise to a wealth of novel physical phenomena[20]. Transition metal oxides (TMOs) are a classic class of correlated electron materials. Owing to the lattice discontinuities and electronic structure reconstructions at the interfaces, novel electronic states such as interfacial magnetic coupling, 2D electron gas and superconductivity have been observed[21-26]. Transition metal nitrides (TMNs) family constitute another important family of correlated electronic materials, attracting considerable interest because of the rich mechanical and physical properties. The lower electronegativity of N and the broader hybridization bandwidth between the transition metal $d$ band and N $p$ band jointly reshape the electronic structure, in clear contrast to TMOs[27]. Benefiting from the advances in epitaxial growth techniques, high quality epitaxial TMNs films exhibit unique interfacial phenomena, such as interfacial magnetism, magnetic proximity effects and proximity ferroelectricity[28-33]. Interfacial engineering in TMOs and TMNs is essential for uncovering exciting phenomena and fabricating next generation electronic devices.

The development of sacrificial-layer techniques enables the study on the vdW interfaces beyond traditional valence or ionic bonded interfaces. The twisted angle-dependent periodical moiré patterns and polar vortices have been experimentally observed in twisted bilayer perovskite TMOs[34-36]. Recent work on twisted bilayer $SrTiO_3$ exhibits lattice-dependent charge disproportionation tied to the local atomic registry[37]. Our previous work shows the moiré modulation on Cuire temperature at a vdW TMOs magnetic interface, suggesting the potential to control magnetic interactions in epitaxial membranes via moiré potential[38]. However, the vdW interlayer coupling and moiré characters in TMNs has not been reported yet, largely due to the difficulty of integrating the TMNs with conventional sacrificial layers. As summarized in Figure 1a, common complex sacrificial oxides such as $Sr_3Al_2O_6$ (or $Sr_4Al_2O_7$) and TMOs families require an oxygen pressure to maintain the stoichiometry[39-57]. In contrast, most TMNs must be grown under high vacuum ( < $10^{-5}$ Pa) and small amount oxygen concentration leads to inevitable oxidation[28, 30, 58-63]. We noticed that the alkaline-earth metal monoxides (MO, M=Ca, Ba, Sr) with rock-salt structure can spontaneously form stoichiometry and structurally stable lattices under high-vacuum condition[64-67]. Their low lattice energy and highly exothermic hydrolysis process confer excellent water solubility, suggesting strong potential for integration with TMNs[68].

In this study, we report the use of (001)-oriented SrTiO$_3$ substrates to grow high-quality, water-soluble BaO thin films, on which epitaxial CrN layers—an antiferromagnetic metal—are subsequently deposited. Freestanding CrN (FS-CrN) membranes are obtained by selectively etching the BaO layer in deionized water. Four-dimensional scanning transmission electron microscopy (4D-STEM) of twisted CrN membranes (tCrN) reveals a periodic, square-like moiré lattice at the interface between two CrN layers. By precisely controlling the twist angle, we identify a $\Sigma = 25$ nearly commensurate moiré lattice in 16.3°-tCrN. Furthermore, for a twist angle of 45°, we experimentally observe localized quasicrystalline order with eightfold rotational symmetry and clear self-similarity. These results demonstrate the feasibility of realizing van der Waals–like interfaces using simple alkaline-earth metal monoxides and establish a basis for tuning interfacial lattice and electronic structures via moiré modulation.

Pulsed laser deposition (PLD) was employed to epitaxially grow CrN/BaO/SrTiO$_3$ heterostructure. A homemade BaO$_2$ target was used for depositing high-quality BaO epitaxial thin films (see methods and fig. S1). After *in-situ* annealing the BaO layer for 30 minutes, a ~16 nm-thick CrN layer was deposited subsequently. During the growth process, the chamber pressure was maintained below $2\times10^{-8}$ Torr to prevent CrN layer from oxidizing. Figure 1b shows the epitaxial relationship between the CrN firm, the BaO sacrificial layer and SrTiO$_3$ substrate. The intrinsic cubic-on-cubic stacking order leads to a large lattice mismatch between BaO ($a_{BaO}$=5.58 Å) and SrTiO$_3$ ($a_{SrTiO3}$=3.905 Å). An in-plane rotation of BaO lattice by 45° maximizes the lattice matching, reducing the mismatch between $a_{BaO}$ and $\sqrt{2}\ a_{STO}$ to ~0.23%[69].

The XRD *2θ-ω* scan in Figure 1c reveals a BaO lattice parameter of 5.56 Å and excellent crystallinity, as evidenced by the narrow rocking curve and pronounced Laue oscillations (figs. S2 and S3). The Phi scan further confirms the 45° in-plane rotation of BaO epitaxially grown on SrTiO$_3$ substrate (fig. S2). The CrN (002) diffraction peak appears at 43.44°, corresponding to a lattice parameter of 4.16 Å, indicating that the CrN layer under a compressive stress when grow on the BaO layer. The CrN layer can be released from substrate by dissolving the entire stack in deionized water, yielding large-area FS-CrN with smooth surfaces (Figure 1c and fig. S4). When the FS-CrN is transferred onto an Al$_2$O$_3$ substrate, the CrN (002) diffraction peak shift to higher angle (~43.81°), corresponding to a contracted lattice parameter of 4.13 Å, suggesting lattice relaxation upon detachment from the BaO layer. The core-level XPS spectra of the FS-CrN are illustrated in Figure 1d and fig. S5. The deconvolution of Cr $2P_{1/2}$ and Cr $2P_{3/2}$ core levels yields single components at 584.7 eV and 575.1eV, consistent with stoichiometry CrN and showing no evidence of metallic Cr or Cr$_2$N phases[70, 71]. The additional peak at 398.3 eV in the N 1*s* spectrum is attributed to unavoidable oxidization during the etching and transfer processes. The release of compressive stress together with the maintaining of stoichiometry demonstrates the feasibility of obtaining high-quality freestanding TMNs using water-soluble alkaline-earth metal monoxides as sacrificial layers.

In general, twisted 2D crystals with hexagonal lattices give rise to rhombus or hexagonal moiré lattices[1, 3, 72]. In contrast, TMOs and TMNs with perovskite or rock-salt structures, which exhibit a 2D square Bravais lattice, can generate a distinct square-like moiré lattice[35, 36]. Twisted bilayer CrN membranes (tCrN) with controlled twisted angles were fabricated using own development NaCl-assistant transfer method (see methods). The advanced four-dimensional scanning transmission electron microscopy (4D-STEM) was employed to elucidate the atomic structure of tCrN (Figure 2a and fig. S6). By varying the defocus depth of 4D-STEM, clear and bright Cr columns are observed in both the top and bottom CrN layers, suggesting the high crystallinity of FS-CrN (Figures 2b and 2c). In contrast, when the focus is set at the interface between two CrN layers, a periodic moiré lattice emerges, as illustrated in Figure 2d. Two distinct sets of diffraction spots, marked by red and blue squares in the Fast Fourier Transform (FFT) image (Figure 2e), correspond to the top and bottom CrN layer, respectively, from which the twisted angle $\alpha=17.8°$ is readily extracted. Figures 2f and 2g show a magnified view of the square-like moiré lattice and the corresponding simulated structure. Red squares denote individual moiré patterns, whereas white arrows indicate the moiré vectors (whose magnitude define the moiré wavelengths), oriented along two mutually perpendicular directions. Note that Nitrogen (Z=7) provides negligible contrast, thus the moiré lattice arises from interference between two Cr sublattices. Moreover, the moiré lattice only appears at the interface region, while the intrinsic lattice periodicity maintained within the FS-CrN, further confirming that the moiré modulation is an interfacial effect (fig. S7).

The atomic line profiles of top and bottom CrN layers are shown in Figure 2h. The measured lattice parameter of FS-CrN is approximately 4.12 Å in both CrN layers and is isotropic within the in-plane orientation. The same lattice parameters in tCrN suggest that both CrN layers are fully relaxed and do not experience significant residual strain. Previous work reported that local atomic registry can lead to strong charge disproportionation of the transition metal atoms in twisted bilayer TMOs[37]. However, in tCrN, no significant changes in the bonding states at the N *K*-edges and Cr-*L* edges are observed, either in single layer region or in moiré lattice region (Figure 2i and fig. S8). This difference may originate from equivalence of the Cr and N sublattices in the rock-salt structure together with the strong Cr-N bonding. The preserved bonding states and relaxed lattice give rise to an "inert" interface, providing an opportunity to reveal the pure vdW moiré modulation in tCrN.

The high-precision displacement stage enabled precise control over the twisted angle. We fabricated a series of tCrN samples with twisted angle $\alpha$ ranging from 5.2° to 28°, and the corresponding high-resolution STEM image are shown in Figures 3a to 3d, respectively. Clear and periodic square-like moiré lattices are observed in all images with the moiré patterns getting denser and the moiré wavelengths monotonically decreasing as the $\alpha$ increase. The corresponding FFT images and simulated atomic structures (Figures 3e to 3h) further highlight the geometry characteristics of square-like moiré lattice in tCrN. We also noted that the local disorder in the moiré lattice can be attributed to the atomic defects or local stress concentration induced during the etching and transfer processes.

The evolutions of the moiré wavelengths of CrN (rock-salt lattice), $La_{0.8}Sr_{0.2}CoO_3$ (LSCO, perovskite lattice) and graphene (hexagonal lattice) are summarized in Figure 3i[38, 73]. For all three lattices, the moiré wavelengths $\lambda$ exhibit an inverse relationship with the twisted angle: $\lambda \sim \frac{a}{2\sin(\frac{\alpha}{2})}$ (where $a$ is the lattice parameter), highlighting the geometric origin of the moiré lattice. The expression indicates that $\lambda$ is solely determined by $a$ when $\alpha$ is fixed, so one would expect $\lambda_{CrN} > \lambda_{LSCO} > \lambda_{Graphene}$ at a given $\alpha$, because $a_{CrN}$=4.12 Å, $a_{LSCO}$=3.83 Å and $a_{Graphene}$=2.46 Å. However, our experiments reveal that the actual trend is $\lambda_{LSCO} > \lambda_{Graphene} > \lambda_{CrN}$, implying that tCrN exhibits the smallest moiré wavelength despite having the largest lattice parameter. The underlying reason is straightforward: in the rock-salt structure, the 2D projection renders all Cr lattice sites equivalent, leading to an effective in-plane lattice parameter $a_{tCrN}$=0.5$a_{CrN}$=2.06 Å.

The unique atomic configuration in tCrN provides an opportunity to investigate the hidden exotic structures in TMNs lattices. By varying the twisted angle, previous work observed both commensurate and incommensurate structures in twisted bilayer system[15, 74]. The commensurability of a twisted bilayer system is concretely manifested by the emergence of a coincidence site lattice (CSL) with translational symmetry[73]. Σ, defined as the ratio of the area of a commensurate moiré supercell to that of the primitive cell, is used to quantify the CSL[75]. We calculate Σ for 2D square and hexagonal Bravais lattices in Figure 4a, with Σ limited to 500 (see methods). Because the CSL is solely determined by the configuration of 2D Bravais lattice, changes in the atomic basis can modify the detailed moiré pattern but do not alter the set of commensurate angles. As a consequence, for (001) orientation of rock-salt and perovskite lattices, Σ and the commensurate angles are identical because both project onto the same 2D square Bravais lattice. Likewise, 2D crystals such as graphene and (111) orientation of 3D rock-salt and perovskite lattices share the same Σ and commensurate angles, as they all correspond to the same 2D hexagonal Bravais lattice. The commensurate angles of a 2D square Bravais lattice form a discrete set that is symmetrical about 45°, reflecting the fourfold rotational symmetry of the intrinsic lattice. Whereas the commensurate angles of a 2D hexagonal Bravais lattice are symmetric about 30°, as confirmed by previous report[17].

Commensurate moiré lattices have been experimentally observed in twisted 2D crystals and TMOs but have not yet been achieved in twisted TMNs[37, 75]. In our experiment, a nearly commensurate 16.3°-tCrN was fabricated, in which a Σ=25 CSL appears at the interface, marked by a white dashed square (Figure 4b and fig. S9). The coincident Cr columns (white circles) appear periodically along the [100] and [010] orientations, and the distance between two coincident Cr columns is approximately 1.1 Å, corresponding to a reconstructed *5 $a_{tCrN}$×5$a_{tCrN}$* CSL. The periodical coincident Cr columns can be viewed as the commensurate rotation faults (CRFs), which represent the commensurate stacking arrangements[73]. We observed two distinct moiré patterns that appear alternately in 16.3°-tCrN, which are marked as type-*I* and type-*II*, respectively (Figure 4c). Type-*I* has four partially overlapped Cr columns located at the center, while type-*II* is characterized by a CRF containing a single coincident Cr column. Type-*I* can transform into type-*II* and form a new CRF via interlayer gliding along [100]/[010] orientations (the gliding vector is indicated by the red arrow), with an ideal gliding distance $d_1 = \frac{1}{5}a_{tCrN} = 42\ pm$. In addition, type-*II* can be mapped onto itself by gliding along [110] orientation over a distance of $d_2 = \sqrt{2}d_1 = 59\ pm$, marked by the blue arrow. Figure 4d shows the gliding translation square which describes the gliding relations among the CSLs in 16.3°-tCrN. The CRFs periodically appear when interlayer gliding occurs along [100]/[010] and [110] orientations, further confirming the transitional symmetry of CSL. Multiple kinds of CRFs are observed in twisted TMDs due to variations in sublattice exchanging, while only one type of CRF appears in tCrN due to its simple structure[75]. The gliding translation square shows a simple and straightforward configuration, which is different from the gliding transition diamond in twisted 2D hexagonal lattices, where different stacking transitions occur between multiple kinds of CRFs[73].

In Figure 4e, we illustrated the measured $d_1$ and $d_2$ values extracted from the STEM image. Note that the spatial resolution of STEM techniques is finite, so the measured values of $d_1$ and $d_2$ are discrete. The fitting values of $d_1$ and $d_2$ are approximately 45 pm and 59 pm, which match well with the ideal value of 42 pm and 59 pm, respectively. Moreover, the interlayer gliding is experimentally observed in the 16.3°-tCrN. Figure 4f shows two CSLs marked by blue and red squares, separated by a transition region with a width of approximately 3Å. The vertical offset $d_{offset}$ of the two CSLs is ~5.2 Å, which is half of the edge length of the Σ=25 CSL ($5a_{tCrN}$). The shift between the two CSLs can be viewed as partially interlayer gliding along the $[0\bar{1}0]$ orientation with a distance of $d_1$; detailed simulations and analysis can be found in fig. S10. We speculate that the coexistence of two CSLs in one local region can be attributed to the energetic favorable commensurate structures[15, 76, 77]. The partial displacement of one CrN layer induced by external perturbations leads to deviation from the original CSL. However, these external perturbations do not induce a global lattice distortion but instead drive the lattice to glide along a specific direction and stabilize in a "neighboring" CSL configuration[75, 76]. Meanwhile, a transition region forms between the two CSLs, acting as a buffer that accommodates lattice distortions and local stress. This exotic phenomenon indicates that robust commensurability can be achieved in twisted TMNs.

Quasicrystal (QC), characterized by lack of translational symmetry while maintaining rotational symmetry, has greatly expanded our understanding of crystallography. Recently, twistronics has provided a route to fabricate the "artificial" 2D QCs beyond conventional rapidly annealed alloys. 2D crystals such as graphene and TMDs can form dodecagonal QC with twelvefold rotational symmetry and exhibit strong interlayer coupling via Umklapp scattering mechanism[17, 19, 78]. Meanwhile, the octagonal QCs with eightfold rotational symmetry have been predicted to exist in the twisted 2D square Bravais lattices[79]. However, it has not been experimentally observed, mainly due to the hurdles in creating high quality freestanding membranes and forming clean vdW interfaces with precisely controlled twisted angles. Figure 5a and fig. S11 show the atomic structure of 45°-tCrN. Unlike the commensurate moiré lattice, 45°-tCrN shows no periodical CRFs, suggesting an incommensurate structure. However, the special moiré configuration shows an eightfold rotational symmetry, which is a key characteristic of an octagonal QC[80]. The atomic structure of 45°-tCrN can be described using an Ammann-Beenker (AB) tiling model, which is a conventional geometry framework for octagonal QCs[80, 81]. Rhombuses and squares with the same edge length serve as the basic units and can tile the entire plane. AB tiling also features self-similarity, meaning that the rhombuses and squares can be inflated by a specific scaling factor while still filling the entire plane. The scaling factor in octagonal QCs is $\sqrt{2}+1$, which is different from that in dodecagonal QCs of $\sqrt{2+\sqrt{3}}$ in twisted 2D crystals, described by the Stampfli tiling model[78].

In order to verify the existence of octagonal QCs, we fabricated a tCrN with the twisted angle precisely set to 45°. The moiré structure at the interface is consistent with our expectation: the actual structure of 45°-tCrN agrees well with the simulation, and the moiré lattice can be spatially mapped onto an AB tiling model, as shown in Figure 5b. The basic AB tiling and its first and second inflations are observed, and all rhombuses and squares at different inflation levels can be assigned to realistic local atomic configuration. In addition, the corresponding FFT image also shows an eightfold rotational symmetry, further confirming the existence of an octagonal QC in 45°-tCrN. Note that our sample is not perfectly aligned to 45° and local stress concentration can lead to distortions of AB tiling. As a result, the observed QC is limited to a confined region and the high-order diffraction spots in the FFT image are suppressed (inset of Figure 5b). We further studied the self-similarity characteristic of the octagonal QC in 45°-tCrN. We statistically analyzed the measured edge lengths ($L_1$, $L_2$ and $L_3$) of rhombuses and squares at different inflation levels (Figures 5c and 5d). The average values of $L_1$, $L_2$ and $L_3$ are 2.48, 6.05 and 14.46 Å, respectively. The evolution of the edge lengths can be described by an exponential function: $L_n = (\sqrt{2} + 1)^{n-1} L_1$, which matches well with the scaling factor in AB tiling (Figure 5e). This phenomenon further confirms the formation of an octagonal QC in 45°-tCrN from both geometric and mathematical perspectives.

Recent work on twisted TMOs has already provided solid evidence that the moiré potential can strongly influence the local atomic registry and electronic structure. Related theoretical calculations suggest that the moiré modulation on electronic structure in twisted TMOs can be attributed to the interfacial lattice reconstruction and orbital hybridization, which are also the origin of emerging phases in strong correlated materials[37, 82]. In this sense, the moiré modulation still acts on traditional order parameters but through a new degree of freedom. The hidden connection between moiré potential and the electronic correlation phenomena remain largely unexplored. From an experiment perspective, the gap between two freestanding membranes—typically ranging from several angstroms to one or two nanometers—can weaken the interfacial coupling[37, 38, 83, 84]. However, certain strong phenomena involving strong electronic correlations, such as interfacial magnetic interaction, proximity superconductivity and ferroelectricity, can remain effective over distances of several to a few tens of nanometers away from the interface[31, 85-96]. We believe that studying on the moiré modulation on above phenomena in twisted TMOs and TMNs can further advance the development of condensed mater physics.

By modifying the twisted angle, we observed intriguing moiré structures in tCrN, including the commensurate/incommensurate moiré lattices and quasicrystalline order. On one hand, the atomic structure is quite different from that of twisted 2D crystals, manifesting as the square-like moiré lattice, a single CRF configuration and eightfold rotational symmetry in the octagonal QC. On the other hand, tCrN also shows similarities to twisted 2D crystals, such as the evolution of moiré wavelength and the self-similarity of the quasicrystal structure. Beyond modulation on lattice structure, our work provides a strong basis for studying the moiré modulation on physical properties of twisted TMNs. As reported in the twisted 2D $CrI_3$, the twisted angle can significantly influence the magnetic ground states and can lead to the coexistence of ferromagnetic state and antiferromagnetic state, depending on the local stacking order[97, 98]. We believe that the moiré effect can likewise be used to further tune the antiferromagnetic order in tCrN. In future work, advanced transfer techniques, such as inducing a Ni capping layer or employing water vapor assistant transfer methods, can be used to fabricate clean and almost damage-free surfaces of freestanding membranes[99, 100]. In addition, developing new theoretical methods will be essential for revealing the novel structures and hidden phenomena at twisted TMNs interfaces.

In summary, we have fabricated twisted bilayer CrN membranes and experimentally observed both commensurate moiré lattices and octagonal quasicrystalline order. At the tCrN interface, we observe a distinctive square-like moiré lattice whose wavelength follows the universal relation $\lambda \sim \frac{a}{2\sin(\frac{\alpha}{2})}$. A $\Sigma = 25$ commensurate moiré lattice, characterized by coincident Cr columns, is realized in 16.3°-tCrN. The interlayer registry can be described by a gliding-translation square, with a concrete example provided by partial gliding along the $[0\bar{1}0]$ direction. In addition, an octagonal quasicrystal with eightfold rotational symmetry is observed in 45°-tCrN, in agreement with earlier theoretical predictions. Multiple inflation levels are identified, with a self-similarity scaling factor of $\sqrt{2}+1$. Our work demonstrates the feasibility of creating vdW–like interfaces in twisted bilayer transition metal nitrides and establishes a platform to explore the reconstruction of antiferromagnetic order and other hidden moiré-related phenomena in tCrN and other twisted TMN interfaces.

**Acknowledgements**
We thank Prof. Jun Kang at Beijing Computational Science Research Center. for fruitful discussions. This work was supported by the Beijing Natural Science Foundation (Grant No. JQ24002 to E.J.G.), the National Natural Science Foundation of China (Grant Nos. U22A20263 to E.J.G. and 12304158 to Q.J.), the CAS Project for Young Scientists in Basic Research (Grant No. YSBR-084 to E.J.G.), the CAS Youth Interdisciplinary Team, the Special Research assistant, the CAS Strategic Priority Research Program (B) (Grant to E.J.G.), the Guangdong Basic and Applied Basic Research Foundation (Grant No. 2022B1515120014 to S.M.Wang), the China Postdoctoral Science Foundation (Grant No. 2022M723353 to Q.J.), and the International Young Scientist Fellowship of IOP-CAS to S.Choi. XPS experiments were performed at IOP-CAS via a user proposal.

**Competing of Interests**
The authors declare no conflict of interest.

**Data Availability**
The datasets generated during and/or analyses during the current study are available from the first author (D.K.R.) and corresponding authors (E.J.G.) on reasonable request.

**Experimental section**

Materials synthesis

The home-made $BaO_2$ target: $BaO_2$ powder (99% purity) is compressed into 1-inch-diameter, 5 mm-thick pellet under a uniaxial pressure of 30 MPa for 20 minutes. The pressed $BaO_2$ target was then sintered at 1200°C in a box furnace for 12 hours. It should be noticed that during cooling process, the $BaO_2$ target must be removed from the furnace when the temperature reaches 150°C-100°C to prevent from deliquescence upon exposure to air.

Epitaxial growth of BaO and CrN thin films: BaO and CrN thin films were grown by PLD using a focused XeCl excimer laser with a duration of 25 ns and a fixed wavelength of 308 nm. A 16 nm-thick BaO film was deposited on a (001)-oriented SrTiO$_3$ substrate at 550°C under high vacuum ( < 2×10$^{-8}$ Torr), with the laser energy set to 0.5 J/cm$^2$. Before deposition of CrN thin film, the BaO layer was kept in high vacuum at 550°C for 15 minutes, which is essential for improving the crystallinity and obtaining a flat BaO surface. Subsequently, the CrN thin film was deposited at 550°C with laser energy increased to 0.68 J/cm$^2$. A 15 nm-thick CrN layer was grown for XRD and XPS characterizations, while a 5 nm CrN was used for fabricating twisted bilayer CrN membranes. When removing the CrN/BaO heterostructure from the high-vacuum chamber, the sample was transferred immediately to the glove box, in order to minimize exposure to ambient air.

Fabrication and transfer process of twisted bilayer CrN structure

Poly(methyl methacrylate) (PMMA, 950 A5) was spin-coated onto the surface of CrN/BaO sample immediately upon removal from the growth chamber of the glove box to protect the fragile 5 nm-thick CrN membrane. The spin speed was set to 2000 rpm, followed by heated at 90°C for 10 minutes to form a supporting layer. A polydimethylsiloxane (PDMS) layer was then laminated onto the PMMA layer, yielding a PDMS/PMMA/CrN/BaO heterostructure. This heterostructure was immersed in deionized water for 15 minutes to ensure complete dissolution of the BaO sacrificial layer. The resulting PDMS/PMMA/CrN stack was released from substrate and transferred to the glove box, where it was left to dry naturally for 4 hours to remove residual moisture. The dried PDMS/PMMA/CrN stack was then mounted onto a glass slide clamped by a high-precision transfer stage. By using an optical microscope, the PDMS/PMMA/CrN stack was aligned and laminated onto the center of a 10×10 mm$^2$ NaCl substrate. The glass slide was subsequently heated to 130 °C to remove the PDMS layer. After cooling down to room temperature, the remaining PMMA/CrN/NaCl stack was immersed carefully in acetone to remove the PMMA layer, leaving a single CrN membranes on the NaCl substrate.

A second PDMS/PMMA/CrN stack was prepared and mounted onto a glass slide following the same process. The glass slide was then rotated with high angular precision to set the twisted angle (±0.2°) between the two CrN membranes. The glass slide drops slowly lowered until the two CrN membranes formed intimate contact. The PDMS and PMMA layers were then removed sequentially, yielding a twisted bilayer CrN structure supported on the NaCl substrate.

Another PMMA layer was spin-coated on twisted bilayer CrN/NaCl, and the sample was gently placed on the surface of deionized water. The NaCl substrate dissolved rapidly and sank to the bottom, while the PMMA/twisted bilayer CrN stack remained floating at the water surface due to the surface tension of PMMA film. Subsequently, the PMMA/twisted bilayer CrN stack was transferred onto a copper grid and allowed to dry naturally. Finally, the PMMA supporting layer was removed in acetone, resulting in a clean twisted bilayer CrN membranes with a precisely controlled twisted angle.

XRD and XPS measurements

X-ray diffraction (XRD) θ-2θ scans, X-ray reflectivity (XRR), and Phi scan were performed using a Panalytical X'Pert³ MRD diffractometer with Cu Kα₁ radiation and equipped a 3D pixel detector. Sample thicknesses were determined by fitting XRR curves using GenX software. X-ray photoelectron spectroscopy (XPS) measurements were performed at the Institute of Physics, Chinese Academy of Sciences. Spectra were collected at Cr 2$p$, N 1$s$ and O 1$s$ core-level peaks at room-temperature, respectively.

STEM observations and analysis

The atomic-scale structural characterization of the thin film was conducted using a spherical aberration-corrected transmission electron microscope JEM NeoARM200 (JEOL Ltd., Tokyo) operated at an accelerating voltage of 200 kV. The electron ptychography data were acquired on a high-speed direct electron detection camera (DECTRIS ARINA) which enables high-quality data acquisition at speeds comparable to conventional scanning transmission electron microscope (STEM) measurements. The iterative reconstruction of electron ptychography datasets was performed using the foldslice algorithm. The sample was imaged along the [001] zone axis of the twisted bilayer CrN structure. Atomic distances between Cr atoms were determined by fitting intensity peaks with Gaussian functions. By adjusting the defocus depth, distinct moiré patterns were observed at the interface between top and bottom CrN layers. Analysis of the STEM data and FFT processing were performed using the Gatan Digital Micrograph software.

Calculation of Σ

In twisted bilayer systems, the commensurate moiré lattices is formed only when the following condition is satisfied: $ma_1^{top} + na_2^{top} = m'a_1^{bottom} + n'a_2^{bottom}$, where $a_1^{top}$ and $a_2^{top}$ are primitive lattice vectors of the top layer and $a_1^{bottom}$ and $a_2^{bottom}$ are primitive lattice vectors of the bottom layer, respectively, and $m$, $n$, $m'$ and $n'$ are integers. This condition defines the CSL and guarantees the transitional symmetry of the commensurate moiré lattice. For the twisted hexagonal lattices, the commensurate moiré lattice emerges at discrete twisted angle $\alpha_h$ satisfying: $\cos\alpha_h = \frac{3p^2+3pq+q^2/2}{3p^2+3pq+q^2}$, and $\Sigma_h = 3p^2 + 3pq + q^2$, where $p$ and $q$ are coprime integers and $\Sigma_h$ denotes the CSL index. In contrast, for the twisted square lattices, the commensurate condition is given by: $\cos\alpha_s = \frac{p^2+q^2}{p^2-q^2}$, and $\Sigma_s = p^2 + q^2$.

By definition, the CSL index Σ characterizes the size of reconstructed moiré lattice. Larger Σ values correspond to high-order commensurate angles, leading to denser distributions of commensurate angles and significantly enlarged moiré supercells.

From an experiment perspective, realizing extremely large commensurate moiré lattices is challenging, as small external perturbations can readily destroy global commensurability. Therefore, we focus on the evolution trend of Σ as a functional of twisted angles for both twisted hexagonal and square lattices, with Σ limited to values not exceeding 500.

For the observed commensurate moiré lattice in figure 4b, the integers $p$=4, $q$=3 yield Σ=25. According to the above expression of commensurate angle for twisted square lattices, the corresponding twisted angle is $\alpha_s$=73.74°. Furthermore, owing to the fourfold rotation symmetry of the intrinsic rock-salt lattice, $\alpha_s$ is equivalent to 16.26°.

**Figures and figure captions**

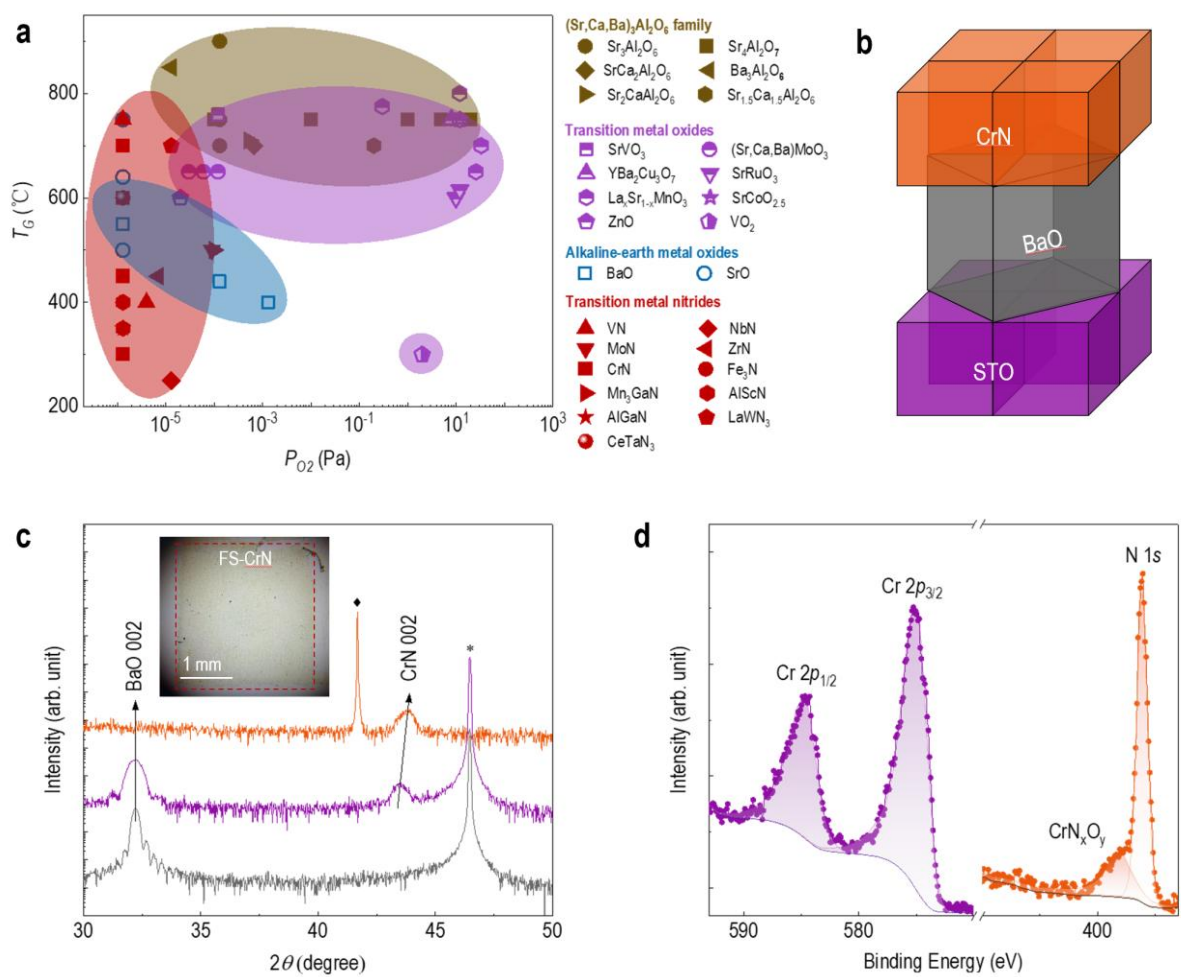

**Figure 1. Synthesis of FS-CrN membranes (FS-CrN).** (**a**) Growth temperature ($T_G$) versus oxygen pressure ($P_{O2}$) for commonly used sacrificial materials and transition metal nitrides. (**b**) Epitaxial relationship of CrN/BaO/SrTiO$_3$ heterostructure with a 45° in-plane rotation of the BaO layer. (**c**) X-ray diffraction (XRD) spectra of BaO/SrTiO$_3$ (gray), CrN/BaO/SrTiO$_3$ (purple) and a FS-CrN membranes transferred onto an Al$_2$O$_3$ substrate (yellow). Inset: optical micrograph of a transferred 15 nm-thick FS-CrN membrane with an area of approximately 2.5×2.5 mm$^2$. (**d**) X-ray photoelectron spectroscopy (XPS) spectrum of an FS-CrN membrane.

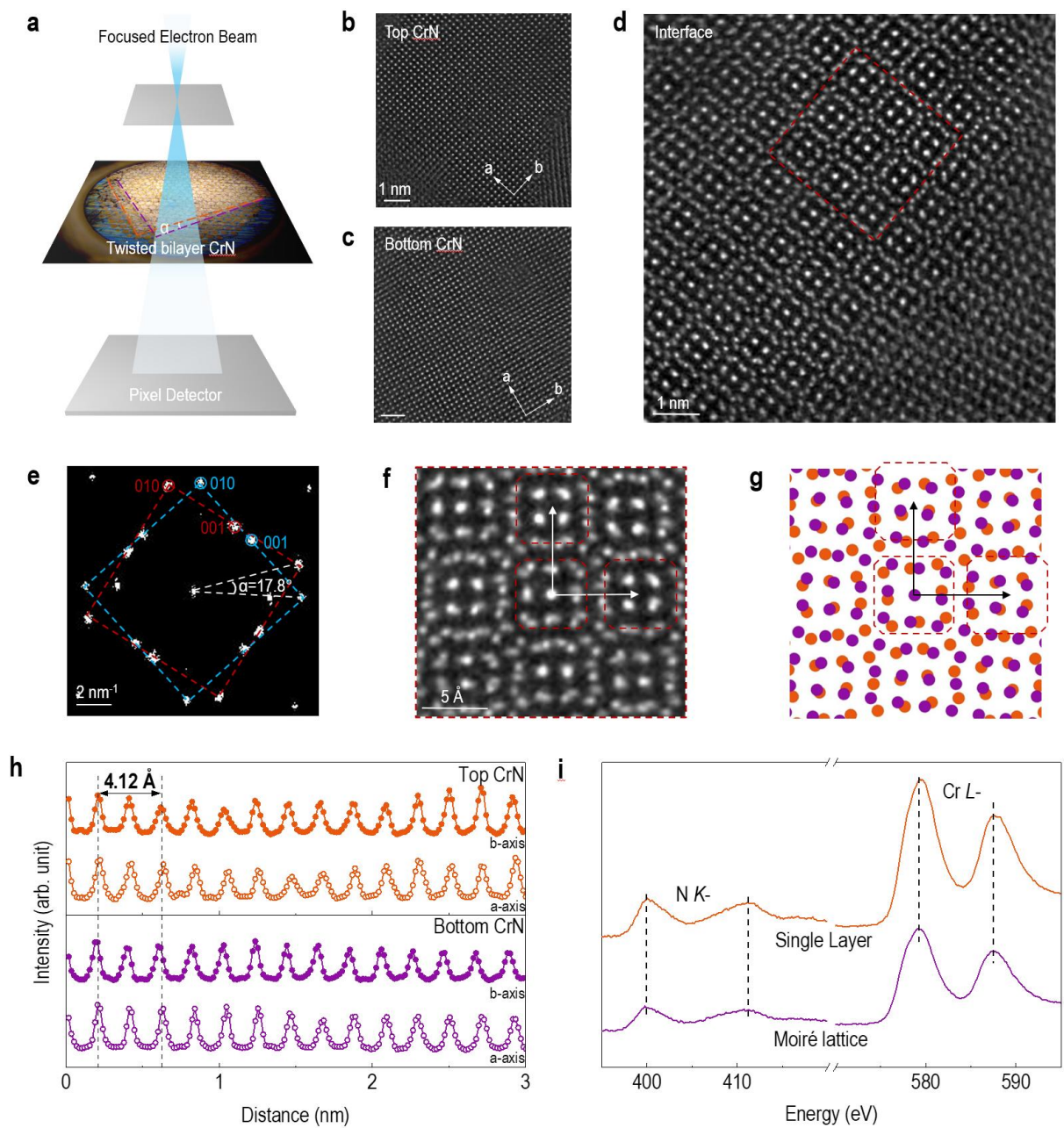

**Figure 2. Imaging the square-like moiré lattice in twisted bilayer CrN membranes (tCrN).** (**a**) Experiment setup for atomic-structure imaging using four-dimensional scanning transmission electron microscopy (4D-STEM). Atomic-resolution STEM images of (**b**) top CrN layer, (**c**) bottom CrN layer and (**d**) interface, acquired by tunning the focal depth. scale bar: 1 nm. (**e**) Fast Fourier Transform (FFT) image of the interface STEM image in (**d**), revealing two distinct sets of diffraction spots (red and blue) corresponding to a relative twisted angle α=17.8°. (**f**) magnified view of the square-like moiré lattice in (**d**), with red squares denote single moiré patterns and white arrows denote the moiré vectors. (**g**) Atomic simulation of 17.8° twisted bilayer CrN, in good agreement with the real-space STEM image in (**f**). (**h**) Intensity profiles obtain from top (**b**) and bottom (**c**) CrN layers. The solid and open circles denote the a- and b-axes, respectively. (**i**) Electron energy-loss spectroscopy (EELS) spectra of single layer region (orange) and the moiré lattice region (purple).

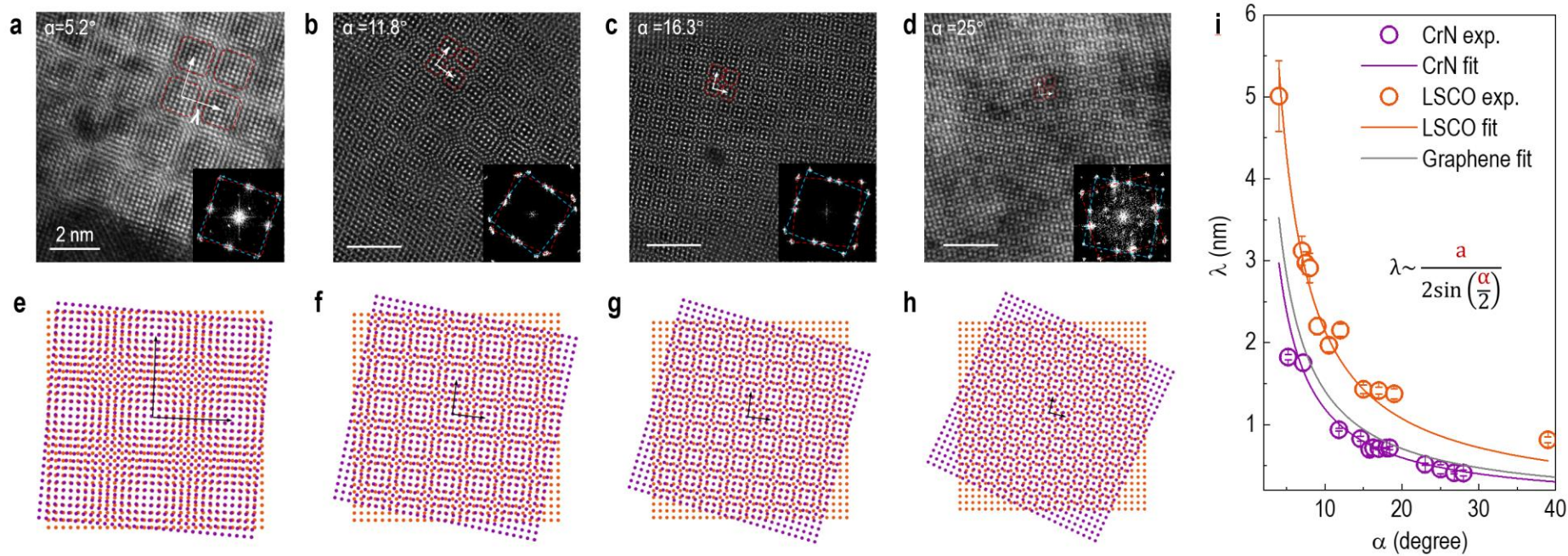

**Figure 3. Evolution of moiré lattice with increasing twisted angle.** (**a**)-(**d**) High-resolution STEM images at twisted angles $\alpha$=5.2°, 11.8°, 16.3° and 25°; insets show the corresponding FFT images. (**e**)-(**h**): Atomic simulations of the moiré lattices corresponding to (**a**)-(**d**). (**i**) Moiré wavelength as a function of twisted angle α for CrN, $La_{0.8}Sr_{0.2}CoO_3$ and graphene with rock-salt, perovskite and hexagonal structures, respectively.

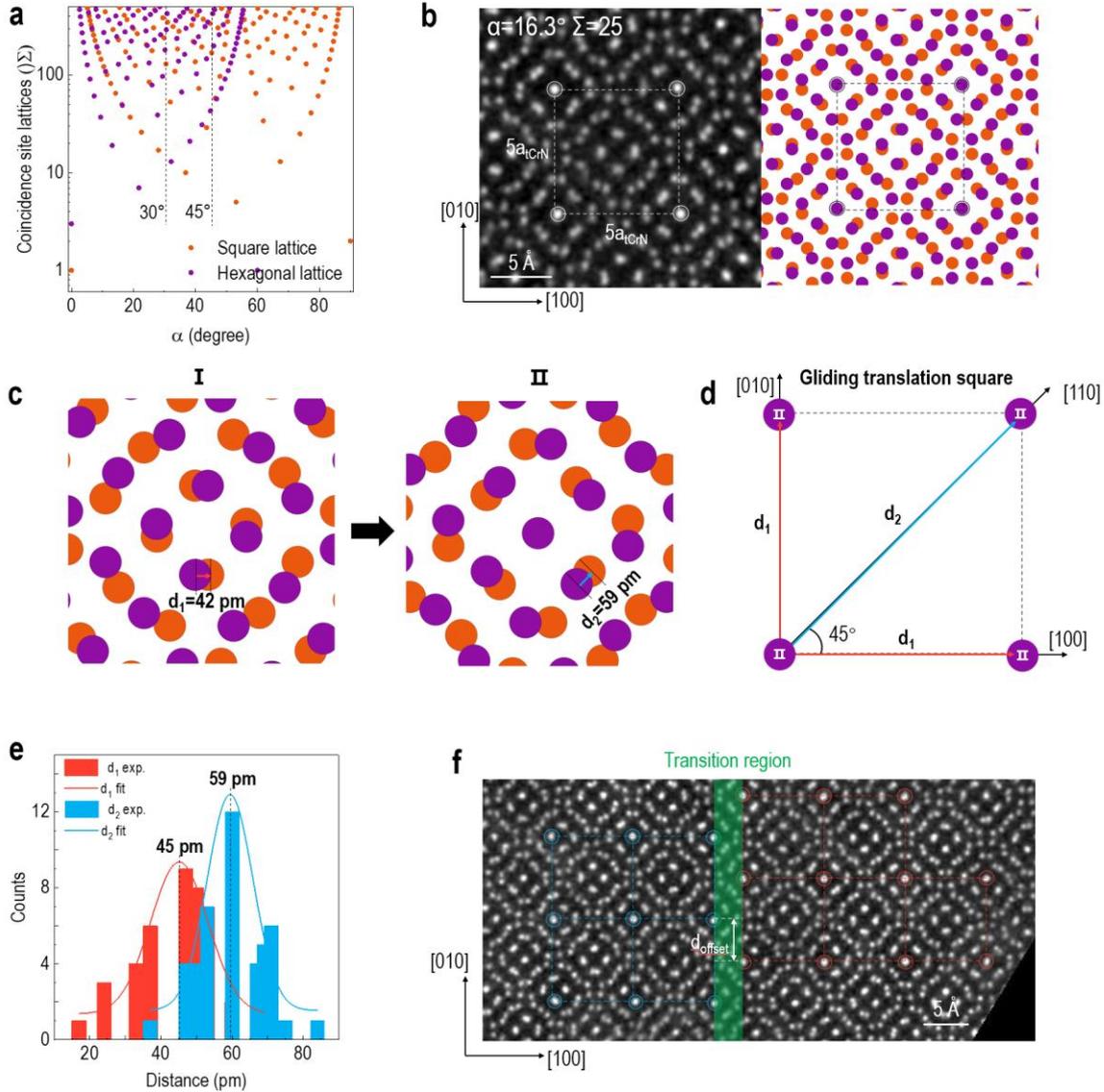

**Figure 4. Nearly commensurate moiré lattice in tCrN**. (**a**) Σ of 2D square and hexagonal Bravais lattices as a function of twisted angle *α*. (**b**) The Σ=25 coincident site lattice (CSL) in 16.3°-tCrN; four white circles mark the coincident Cr columns from the top and bottom CrN layers. (**c**) Type-*I* and type-*II* moiré patterns serving as the basic units in Σ=25 CSL; the type-*II* corresponds to a commensurate rotational fault (CRF). *I-II* and *II-II* transitions can occur along the orientations indicated by red and blue arrows, respectively. (**d**) The gliding translation square illustrates the gliding relation among the CSLs. (**e**) Measured gliding distances for *II-II* transitions along the [110] orientation and *I-II* transitions along the [100]/[010] orientations, extracted from STEM image in (**b**). (**f**) Coexistence of two Σ=25 CSLs (marked by blue and red squares) separated by a transition region (green) in a 16.3°-tCrN.

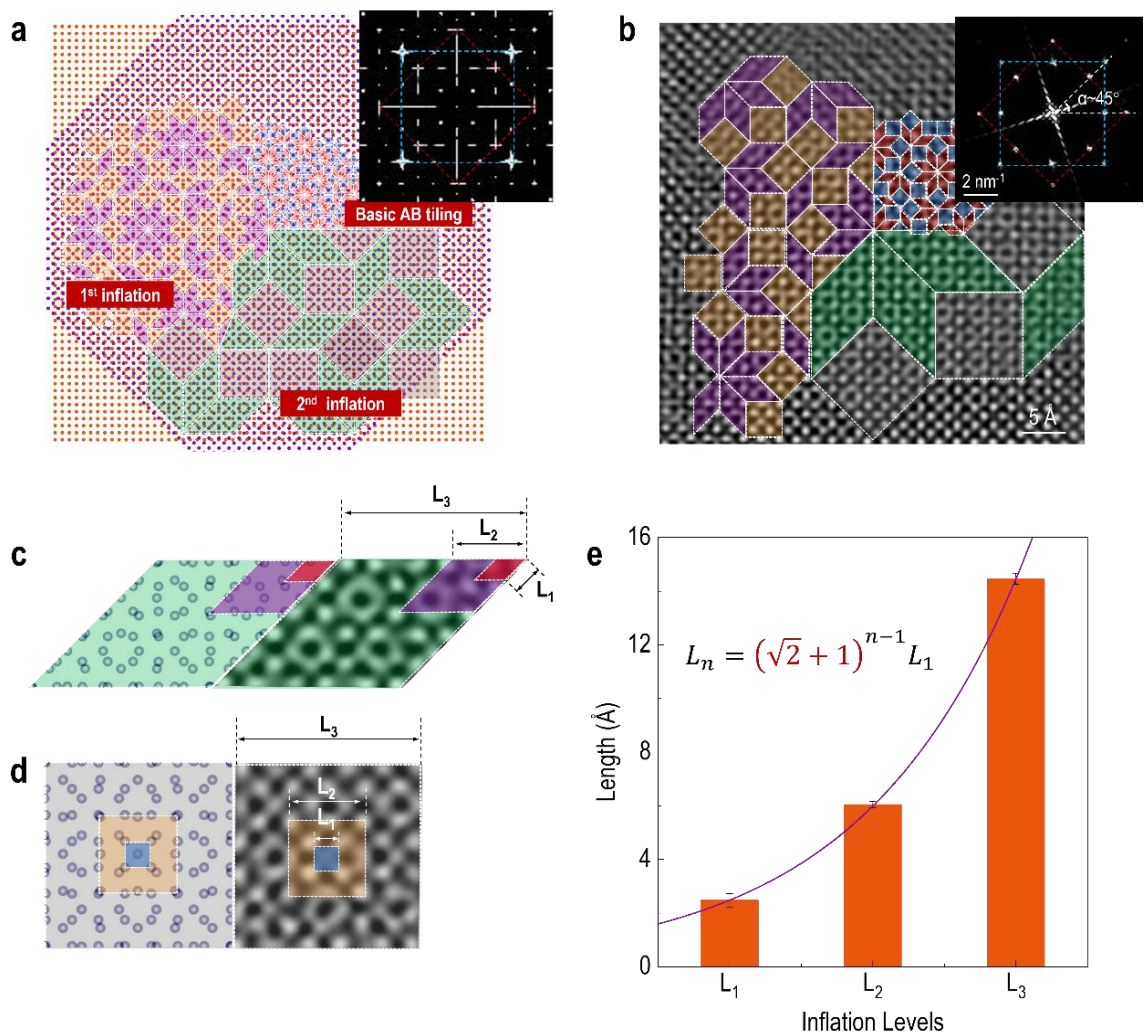

**Figure 5. Localized quasicrystal order with eightfold rotational symmetry in 45°-tCrN.** (**a**) Atomic simulation and (**b**) real-space STEM image of 45°-tCrN, both filled with Ammann-Beenker (AB) tiling. Color-coded rhombuses and squares indicates the basic AB tiling (red and blue), the first (purple and yellow) and second (green and gray) inflation levels with a scale factor of $\sqrt{2}+1$. (**c**) Rhombuses in AB tiling: atomic simulation (left) and corresponding STEM image (right) with $L_1$, $L_2$ and $L_3$ are the edge lengths of basic AB tiling, first- and second-level inflations, respectively. (**d**) Squares in AB tiling: atomic simulation (left) and corresponding STEM image (right) with $L_1$, $L_2$ and $L_3$ defined as in (**c**). (**e**) Statistics of edge lengths extracted from the real-space STEM image in (**b**) for each inflation levels; $L_1$, $L_2$ and $L_3$ increase monotonically by a factor of $\sqrt{2}+1$.